\def\year{2018}\relax
\documentclass[letterpaper]{article} 

\usepackage{aaai18}  
\usepackage{times}  
\usepackage{helvet}  
\usepackage{courier}  
\usepackage{url}  
\usepackage{graphicx} 
\usepackage{alg}
\usepackage[ampersand]{easylist}
\begin{document}
%
\title{Crowdsourcing for Reminiscence Chatbot Design 
}

\author{Svetlana Nikitina \\
University of Trento and\\
Tomsk Polytechnic University\\
svetlana.nikitina@unitn.it
\And  \hspace{0.2cm} Florian Daniel\\
Politecnico di Milano, DEIB\\
florian.daniel@polimi.it
\And  \hspace{0.2cm} Marcos Baez, Fabio Casati \\
University of Trento and\\
Tomsk Polytechnic University\\
baez@disi.unitn.it, fabio.casati@unitn.it 
\And \hspace{0.2cm} Georgy Kopanitsa\\
Tomsk Polytechnic University \\
kopanitsa@tpu.ru
}

\maketitle
  \begin{abstract}
In this work-in-progress paper we discuss the challenges in identifying effective and scalable crowd-based strategies for designing content, conversation logic, and meaningful metrics for a reminiscence chatbot targeted at older adults. We formalize the problem and outline the main research questions that drive the research agenda in chatbot design for reminiscence and for relational agents for older adults in general.
 \end{abstract}



\section{Context \& Objectives}




\textit{Reminiscence} is the process of collecting and recalling past memories through pictures, stories and other \textit{mementos} \cite{webster2007reminiscence}. The practice of reminiscence has well documented benefits on social, mental and emotional wellbeing \cite{subramaniam2012impact,huldtgren2015probing}, making it a very desirable practice, especially for older adults. Research on technology-mediated reminiscence has advanced our understanding into how to effectively support this process, but has reached a limit in terms of the approaches to support more engaging reminiscence sessions,  effectively elicit information about the person, and extend the practice of reminiscence to those with less opportunities for face to face interactions. 

In our previous work \cite{nikitina2018smart} we made a case for conversational agents in this domain, and proposed the concept of a smart conversational agent that can drive \textit{personal} and \textit{social} 
reminiscence sessions with older adults in a way that is engaging and fun, while effectively collecting and organising memories and stories. 
The idea of conversational agents for older adults is not new, and they have been explored to support a wide variety of activities and everyday tasks
\cite{tsiourti2016virtual,vardoulakis2012designing,hanke2016daily,tsiourti2016cameli}, to act as social companions \cite{ring2013addressing,ring2015social,demiris2016evaluation} and even to engage older adults in reminiscence sessions \cite{fuketa2013agent}. 

While these works give us valuable insights into the opportunities of using conversational agents as an instrument to support reminiscence sessions, they also show us how limited our knowledge is in terms of effective strategies 
to maintain dialogs with older adults. Success stories are mostly limited to Wizard of Oz evaluations \cite{schlogl2014wizard}, in which system functionality is partially emulated by a human operator, or based on fully human-operated agents. The few attempts at autonomous agents highlight issues with the mismatch between user expectations and the actual social capabilities of the agents \cite{tsiourti2016virtual}, general challenges with designing conversations suitable to the target population \cite{yaghoubzadeh2015adaptive}, and challenges with engaging older adults in question-based interactions in particular \cite{fuketa2013agent}. 

In this position paper we aim at identifying effective and scalable crowd-based strategies for designing content, conversation rules, and meaningful metrics for a reminiscence chatbot targeted at older adults. We build on the concept introduced in \cite{nikitina2018smart} and identify \textit{where} and \textit{how} crowdsourcing can help design and maintain of an agent-mediated reminiscence process, while addressing the specific challenges posed by the target population.









\section{Reminiscence Chatbot}

\begin{figure}
\centering
\includegraphics[width=.9\columnwidth ]{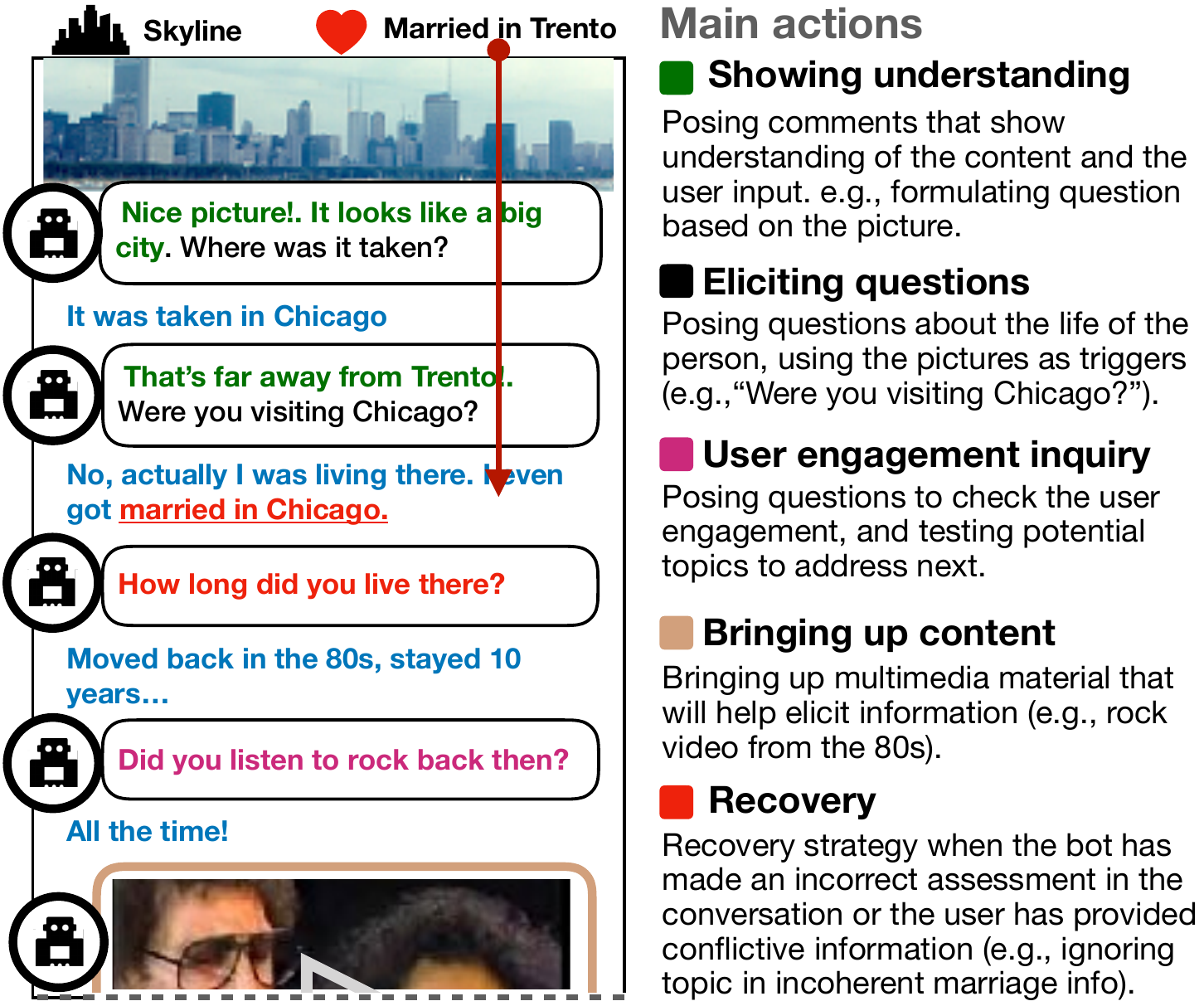}
\caption{Example reminiscence session with bot actions}
\label{figure:concept}
\end{figure}

The envisioned chatbot is based on the idea of automatically guiding older adults through multimedia reminiscence sessions \cite{nikitina2018smart}. It has the dual purpose of i) collecting and organising memories and profile information, and ii)  engaging older adults in conversations that are stimulating and fun. 
In Figure \ref{figure:concept} we show an example conversation and related main actions. 

The example starts from the subject (the elder) providing a memory in the form of a picture. In response, the chatbot poses a contextual question. In order to do so, it must be able to \emph{understand the theme of the picture} (big city) and to \emph{extract and understand information} from pictures and text. In order to keep the conversation natural, it must further be able to \emph{reference related conversation topics} (the city of Trento) and, in order to show empathy, it must be able to \emph{sense the feelings of the subject} as the conversation evolves (e.g., it looks like the subject likes rock music, so it could be an idea to talk about that for some time). It would also be good if the bot be able to \emph{sense the presence of peers} (e.g., family members or moderators helping with the chat). All this information helps the bot \emph{decide on appropriate next actions} taking into account possible \emph{conversational goals} (e.g., elicit basic user profile data). Among the most complex decisions to be taken is deciding if and when to \emph{change context} in a conversation (e.g., to make the elder laugh).

All these requirements are particularly challenging since special attention must be paid to the subject's abilities and limitations \cite{nurgalieva2017designing,hawthorn2000possible}. For instance, it is hard to \emph{cope with user-initiated context switches} or to \emph{keep knowledge about subjects coherent} due to cognitive decline associated with age \cite{park2003systematic}. Coping with these challenges is difficult even for humans \cite{miron2017young}.
\smallskip
In the long term, our goal is to develop a crowd-powered chatbot that implements the necessary conversational logic, sensibility and tricks to engage older adults in pleasant and satisfactory reminiscence sessions. The crowd should not be involved in direct interactions with the elderly (like in some real-time crowdsourcing approaches studied in literature \cite{lopez2016lifeline,ring2015social}), nor should it be used just to train black-box AI algorithms. The idea is to involve the crowd to elicit and represent reminiscence-specific conversation knowledge explicitly in the form of some dedicated model, in order to be able to actively steer the conversation into specific directions (e.g., to elicit health issues or family memories). In this paper, we focus on an intermediate set of research objectives: identifying (i) how to \textit{model} the conversational knowledge the chatbot may rely on and (ii) how to use the crowd to \textit{learn and evaluate} the model.




\section{Crowd-Supported Chatbot Design}

\subsection{Conversational Model Representation}

Conceptually, a simple model we can imagine for a chatbot is a state machine $(S,A,\delta, \pi, F)$, where $S$ denotes the states (a state includes the information on the subject and the conversation history), $F$ denotes the final states, $A$ is the set of (conversational) actions, $\delta$ is a state transition function (our conversational policy), $\pi : S\times A \rightarrow \{(s,p)\}$ associating to each state and action a set $\{(s,p)\}$ of possible target states $s$ and the probability $p$ with which that action should be chosen (to model that conversations are not deterministic). 

In practice however the state space is infinite and the possible conversations are also infinite so this FSM is not the right model. 
An alternative model is based on Event-Condition-Action (ECA) rules, where the event for example is the sentence by the subject (the elder) and the condition is some expression over what we know about the subject as well as past events. This has however the same limitations just discussed.

We observe that what we really want to have is a definition of the domain and range of the policy function $\pi$ so that we can learn a useful policy that can be applied to real life conversations.
On the action side (the range), we approach the problem by clustering similar actions along several dimensions, such as i) the type of actions (ask information, make a comment, show interesting content) and ii) the topic of conversation (talk about the picture you are showing, or about childhood, or about hobbies).
Given the action type and topic, there are many actual conversations and utterances, but at this level we are focused on learning types and topics rather than conducting an interaction within a topic or paraphrasing sentences.

In terms of the domain a policy is defined on, what we wish to have is a description of the characteristics of the state (or event and condition) to which the policy applies. 
For example, the crowd may tell us that after they learn the date of birth, they show newspaper covers of that year, or famous people born the same day, or songs that where popular when the subject was very young.
In this case the trigger of the action is the last conversation element where the subject is notifying the state of birth (or, in terms of events, it is the event of the system, somehow, coming to know the date of birth of the person).   

The challenge here is therefore to understand what is the reasoning of crowd workers when they decide to take actions, and based on this reasoning identify the classes of state and event information we need to attach policies to.

\subsection{Crowdsourcing tasks}

The counterpart of the model is the learning process, which has to do with how to design and process the results of crowdsourcing tasks. 
The objective we have in seeking the proper task designs are the following: (i) identifying \emph{action types and topics} (unless we want to fixe them a-priori), (ii) identifying \emph{when} (based on which state or trigger) a person changes topic or shows specific content, and (iii) identifying \emph{why} (based on which state or trigger) the agent initiates a conversation on a topic. 

To do this, we envision crowdsourcing tasks that aim at (i) exploring possible conversations (these can be Wizard of Oz simulations), (ii) reflecting over previous conversations by the same worker or other workers to derive the ``rules" that made the worker take a certain course of action, and (iii) aggregating these ``rules" into a smaller coherent set that reveals the characteristics that the policy model should have.

For example, the crowd may reveal that they change topic whenever they sense that the person is sad talking about the current topic. This would tell us that an important component of the policy domain is the perceived emotional state, something that therefore the agent should try to detect, and that change in this emotional state should be a trigger to either continue or change topic.

\smallskip

We thus focus on the following research question (\textbf{RQ}): \textit{Which crowd-based strategies can help elicit effective conversation logic for conversations (reminiscence sessions) targeting older adults, and how?}


Conversational logic includes understanding of: 
composition of Dialog State, when and how the State has to be changed, and what are the most important variables that affect the state. 
That is, given:

\begin{itemize}
\item the set of States $S=\{S_1,S_2,...S_m\}$, where S is the state of the conversation that consists of multiple features (such as user profile info, dialog history, sentiments);
\item the set of possible Goals in the conversation $G=\{G_1,G_2,...G_n\}$, where G is the current goal aimed at (e.g., elicit information, tell a joke, show engagement content); and
\item the set of Actions
$A=\{A_1,A_2,...A_n\}$, A being the chatbot action performed, which changes the state and satisfies the current goal (e.g ask question to elicit info);
\end{itemize}

\noindent 
the aim is to:

\begin{itemize}
\item identify the composition of current State; and  
\item identify the \textit{policy}, i.e., which Action to take given current state S and the Goals G 
\item such that 
$$\pi({G,S}) \rightarrow S'$$ 

where Policy $\pi$ is a rule that defines the transition from state S to state $S'$ and depends on the \textit{Current State} S and current \textit{Goals} G of the conversation.
\end{itemize}

The research question is actually of more general nature, and the resulting approach can be applied to any social chatbot. To us, reminiscence is an application domain we have experience with and we want to contribute to.

\subsection{Success Metrics}

Different metrics have been proposed for evaluating the quality of conversations with dialog agents,
such as: i) user engagement \cite{alessandra2017roving,fitzpatrick2017delivering}, ii) task completion \cite{huang2015guardian}, iii) conversation quality: including dialog consistency and memory of past events \cite{lasecki2013chorus}, iv) human-like communication \cite{kopp2005conversational}. The approach to evaluation -- and therefore the choice of metrics -- is based on the aim of the agent: having an engaging chat or performing a specific task (e.g., booking a flight).
%
In our case, the reminiscence chatbot is a combination of conversational and task-based agent, as it aims at both having an engaging conversation with the user and collecting information while doing so. Therefore, we consider metrics for both types of agents, including:
%
i) engagement (as subjective measure); 
ii) number of turns of conversation made before it drops;
iii) times conversation drops overall;
iv) domain-specific metrics like
the amount of content which the user has provided during one conversation session (amount of pictures uploaded, amount of data attributes filled about a relevant person), and other task-completion metrics.

\section{Related work}
Crowdsourcing has been used to support all aspects of chatbot design, from holding direct conversations with final users, to supporting conversation design -- the latter being the family of approaches under which we position our work.  
Prior work on crowdsourcing has addressed the \textit{bootstrapping challenge}, investigating strategies to create dialog datasets to train algorithms~\cite{takahashi2017two,lin2016web}, infer conversation templates~\cite{mitchell2014crowdsourcing} or declarative conversation models~\cite{negi2009automatically}.
It has also been explored to \textit{enrich conversation dialogs} to provide meaning and context, by annotating dialogs with semantics and labels with, for example, polarity and appropriateness~\cite{lin2016web}, extracting entities \cite{huang2016crowd}, as well as providing additional utterances for more natural conversations (paraphrasing) \cite{jiang2017understanding}. Other approaches incorporate the crowd in the \textit{evaluation of chatbot quality}, making sure crowd contributions are valid and safe \cite{chkroun2018did,huang2016there} and even allowing users to train chatbots directly \cite{chkroun2018did}. Acknowledging that chatbot conversations are not perfect, some approaches explore strategies to \textit{escalate conversation decisions} to the crowd in cases where the chatbot is not able to interpret or serve the user request \cite{behera2016chappie}.

The above highlight the potential of crowdsourcing for designing chatbots. We take these approaches as the starting point for exploring the specific challenges of designing and maintaining a reminiscence bot. Previous work in this domain -- though valuable in insights -- has been limited to human-operated chatbots and Wizard of Oz evaluations, highlighting the complexity of chatbot design in general and in particular for our target population \cite{tsiourti2016virtual,fuketa2013agent,yaghoubzadeh2015adaptive}.

\section{Ongoing and Future Work}


Next, we are going to define concrete crowdsoursing strategies to elicit the nature of the states, goals and actions that will give structure to the model. Then, we will focus on tasks to fill the model with data and on algorithms to effectively aggregate and apply the elicited knowledge.


\section{Acknowledgments}
This work has received funding from the EU Horizon 2020 Marie Sk\l{}odowska-Curie grant agreement No 690962. It was also supported by the project ``Evaluation and enhancement of social, economic and emotional wellbeing of older adults" under the agreement No.14.Z50.31.0029, Tomsk Polytechnic University.
\newpage

\bibliographystyle{aaai}
\bibliography{chatbots,reminiscence}

\end{document}